# Tuning the Gate Opening Pressure of Metal Organic Frameworks (MOFs) for the Selective Separation of Hydrocarbons

Nour Nijem[‡,&], Haohan Wu[‡,^], Pieremanuele Canepa[#], Anne Marti[%] , Kenneth J. Balkus Jr.[%], Timo Thonhauser[#], Jing Li[*,^] and Yves J. Chabal[*,&]

[&] Department of Materials Science and Engineering, University of Texas at Dallas, Richardson, TX, 75080, USA,

[^] Department of Chemistry and Chemical Biology, Rutgers University, NJ 08854, USA

[#] Department of Wake Forest University1834 Wake Forest Road, Winston-Salem, NC 27109, USA,

[%] Department of Chemistry, University of Texas at Dallas, Richardson, TX 75080 (USA)

*Supporting Information Placeholder*

**ABSTRACT:** Separation of hydrocarbons is one of the most energy demanding processes. The need to develop materials for the selective adsorption of hydrocarbons, under reasonable conditions, is therefore of paramount importance. This work unveils unexpected hydrocarbon selectivity in a flexible Metal Organic Framework (MOF), based on differences in their gate opening pressure. We show selectivity dependence on both chain length and specific framework-gas interaction. Combining Raman spectroscopy and theoretical van der Waals Density Functional (vdW-DF) calculations, the separation mechanisms governing this unexpected gate opening behavior are revealed.

Metal Organic Frameworks (MOFs) are being considered extensively for their applications in a variety of fields, ranging from gas storage and separation to catalysis, sensing, and drug delivery.[1] Their potential for gas separation and storage stems from their large surface areas and tailorable structures, making their properties readily tunable.[2] Flexible frameworks are particularly attractive for the selective adsorption of gases.[3] Their structural responses to a specific adsorbate, and the possibility to vary the pressures at which different adsorbates are incorporated into the framework have generated much interest in these materials.[3a-d,4]

The selective adsorption of acetylene over methane, ethylene and $CO_2$ by MOFs, for instance, is of practical interest for the separation of gas mixtures in numerous industrial applications.[1c,1e,5] Acetylene is principally derived from the cracking of natural gas. Separation of acetylene from methane is essential for obtaining a Grade A purity for organic synthesis.[5b,6] Separation of light olefins and paraffins is one of the most energy intensive processes, especially when the molecules are close in size.[7] Furthermore, the purification of ethylene by removing the 1% acetylene is essential in producing high quality polymers.[8] Finally, the detection and removal of acetylene in hydrogen is interesting for transformers and reactors.[9]

To date, there have been only a few reports on adsorption-based separation of light hydrocarbon isomers of C2-C4 components using MOFs as adsorbents.[5a,10] Among these studies are separation of C3 propylene and propane by ZIF-8 and TO type MOFs via kinetic mechanism, separation of acetylene and ethylene via gate opening selective adsorption of acetylene in a mixed-metal-organic framework (M'MOF) albeit at 195 K,[11] and separation of ethane and ethylene via gate opening selective adsorption of ethane in ZIF-7.[4a] We present here the first example where separations of C1-C4 paraffins and both C2 isomer pairs ($C_2H_2$-$C_2H_4$ and $C_2H_4$-$C_2H_6$) may be achieved by a *single* MOF *at room temperature* based on the gas-induced structural change and the resulting differences in gate opening pressure. Importantly, combining Raman spectroscopic and vdW-DF methods, we provide a *molecular level understanding* of the structural change associated with the unique gate-opening and stepped isotherms. $Zn_2(bpdc)_2(bpee)$ (bpdc = 4,4'-biphenyldicarboxylate; bpee = 1,2-bipyriylethylene), also known as RPM3-Zn, has a notable stepped room-temperature $CO_2$ adsorption-desorption isotherm with no hysteresis.[12] This is believed to originate from the high structural flexibility of its framework and the nature of the framework-$CO_2$ interaction, as spectroscopically confirmed.[13]

Here, the interactions of light paraffins, olefins, and acetylene within this flexible structure are examined. A pronounced gate opening behavior, followed by stepwise isotherms with a strong hysteresis is observed for both olefins and paraffins, with a clear dependence of the gate opening pressure on the chain length (Figure 1).

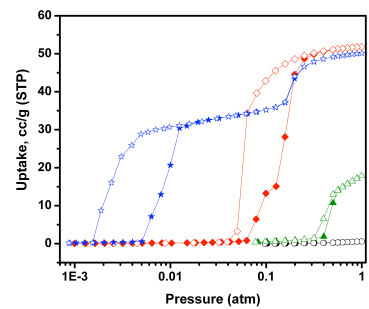

Figure 1. Adsorption-desorption isotherms of short alkanes in RPM3-Zn at room temperature (298K) plotted as a function of relative pressure. Filled and open symbols represent adsorption and desorption branch, respectively. Color schemes: black circles: $CH_4$, green triangles: $C_2H_6$, red diamonds: $C_3H_8$, Blue asterisks: $C_4H_{10}$.

Such stepwise adsorption isotherm is very similar to $N_2$ adsorption at 77K in a flexible framework Co(BDP)·2DEF[14] (BDP = 1,4-benzenedipyrazolate; DEF = N,N'-diethylformamide), for which an in-situ powder X-ray diffraction experiment was performed to analyse the structure transformation during adsorption. It is believed that the molecular kinetic diameter associated with diffusivity and hydrogen bonding between hydrocarbon molecules and MOF framework play a crucial role on such abnormal behavior.[1e,15] It is interesting to note that the unexpected gate opening pressures are lower for acetylene and ethane than for ethylene, while propylene and propane have the same gate opening pressure (see Supplementary Information, Figures S1 and S2).

To determine the gate opening mechanism, we performed Raman spectroscopy measurements in conjunction with first-principles density functional theory (DFT) calculations, using the recently developed van der Waals density functional vdW-DF method.[16] Raman measurements for ethane adsorption in RPM3-Zn (Figure 2) reveal several significant changes in the fundamentals modes of the host sorbent ligands associated with the presence of ethane: 1) a decrease of the C=O stretch mode intensity at 1650 cm$^{-1}$, and 2) a ~2 cm$^{-1}$ red shift of the C=C mode at 1643 cm$^{-1}$, of the asymmetric stretch mode of the C-O mode and of the in-plane pyridine stretch both at 1611 cm$^{-1}$. Furthermore, in the lower frequency range, a weak mode at 991 cm$^{-1}$, assigned to the C-C stretch mode of adsorbed ethane, is red shifted by ~-4 cm$^{-1}$ from the unperturbed C-C stretch mode in free ethane. A new mode at 886 cm$^{-1}$ emerges and is identified as the out-of-plane deformation mode of the pyridine ring. An intensity increase of the in-plane deformation mode of the pyridine ring (green labels) is also observed, along with a loss of intensity of the mode at 1286 cm$^{-1}$, assigned to the inter-ring C-C stretch mode in the bpdc ligand (red label).

ethane and the non-coordinated C=O of the framework. The initial adsorption configurations of the molecules are deduced from the Raman data and accepted guest-host interactions (see Figure S13 in the supplementary material). For the calculations, the structure was fully relaxed to take into account structural changes that occur due to adsorption. Previous studies in MOFs have shown that H-bonding is a way to achieve preferential acetylene adsorption and is also a mechanism to induce a gate opening behavior for the adsorption of molecules such as water and methanol.[4b,17] However, it is the first time that the hydrogen bond strength is shown to be a factor affecting the gate opening *pressure* as shown in this work.

The calculations also show that ethane incorporation into the structure causes the dihedral angle between the two rings of the bpdc ligand to decrease by Δφ= -2.0 degree (Figure 3b) from its original position (φ= 26.8 degree). This change in angle is consistent with the loss of intensity of the inter-ring C-C mode (i.e. the bpdc linker becomes flatter). Furthermore our calculations confirm that small perturbations of the dihedral angle φ, in the fully relaxed RPM3-Zn, do not introduce intolerable structural strains, hence supporting the existence of the gate opening mechanism (see Figure S12 in the supplementary information). The ethane interaction and position explains the disappearance of the C=O mode, i.e. its dramatic decrease in Raman activity. The activation of the out-of-plane deformation modes at 886 cm$^{-1}$ is the result of the C=O change in position close to the pyridine ring. The red shift in the C=C and monodentate C-O stretch modes is consistent with structural rearrangement leading to an opening of the structure.

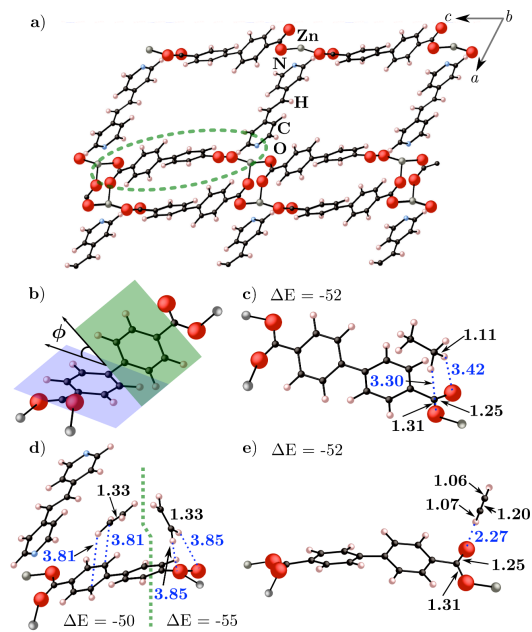

Figure 3. Side view of the RPM3-Zn structure (a), dihedral angle in the bpdc ligand (b), local fragments of ethane (c), ethylene (d), showing the interaction with the inter-ring C-C (left) and with the C=O (right), and acetylene (e), respectively adsorbed in the RPM3-Zn. Bond lengths in blue are in Å and ΔE in kJ mol-1. For ethylene two possible adsorption sites are presented.

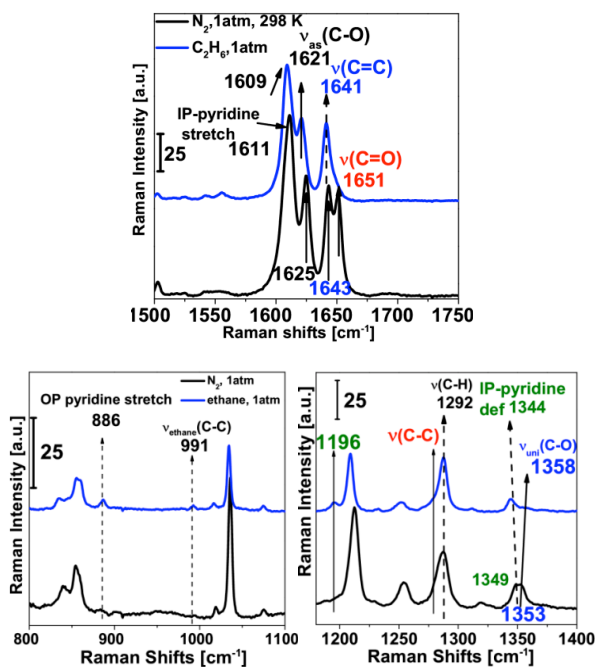

Figure 2. Raman spectra of RPM3-Zn in 1 atm N2 (black spectrum) and after introducing 1 atm of ethane at room temperature (298 K) (blue spectrum) the top and the bottom panels show the spectra in different frequency range.

The vdW-DF calculations summarized in Figure 3c show that ethane interacts non-specifically through its $CH_3$ groups with the framework monodentate carboxylate group present in the 2D layers (adsorption energy (ΔE) ~ -52 kJ mol$^{-1}$), suggesting the formation of a weak hydrogen bond between the $CH_3$ of the

To determine that the structure indeed changes due to adsorption of hydrocarbons, we performed *ex-situ* X-ray diffraction measurements on RPM3-Zn after adsorption of a hydrocarbon that is a liquid at ambient conditions, such as 1-hexene (section 3, Figure S3 in supporting information). The RPM3-Zn is seen to exhibit reversible structural changes as a function of adsorption and desorption of 1-hexene. Raman measurements (Figure S4) for 1-hexene adsorption show similar effects to those observed for ethane adsorption (Figure 2). Moreover, it can be noted from the isotherm data (Figure 1) that the gate opening pressure point requires a number of molecules to interact with the two non-coordinated C=O of the bpdc ligands in the 2D layers (Figure 3a).

The interaction of ethylene in RPM3-Zn was examined, in comparison to ethane. As shown in section 4 of the supplementary information (Figure S5), the Raman spectra of ethylene adsorption suggest a different adsorption mechanism than that observed for ethane. Indeed, the vdW-DF calculations show that there is a notable H-bonding for ethylene, very similar to the case of ethane, between the $CH_2$ of the ethylene and the C=O bond of the ligand with a ΔE of -55 kJ mol$^{-1}$ (Figure 3d). Additionally, an interaction of the C=C bond of ethylene with the C-C inter-ring of the bpdc ligand is also possible, with a ΔE = 50 kJ mol$^{-1}$ (Figure 3d), increasing the dihedral angle between the two benzene rings (Δφ = 1.5 degree). These results account for the Raman observations of a ~-6 cm$^{-1}$ red shift of the C-C inter-ring mode, and a ~3 cm$^{-1}$ blue shift of the C=C stretch mode in the bpee ligand. This interaction presents a competitive alternative binding site with similar binding strength to that of the hydrogen bonding. Therefore, more ethylene molecules are needed to satisfy the hydrogen bonding at both ends close to the monodentate C=O of the 2D layer, thus requiring a larger pressure to open than with ethane, because some of the molecules are attracted at this secondary site.

To investigate the role of the presence of $CH_3$ groups on the interaction in unsaturated hydrocarbons, such as propylene, we compared the interaction of propane and propylene in RPM3-Zn. Isotherm measurements performed at room temperature (Figure S2 in supplementary information) show that the pore opening pressure is similar, with a slight difference at higher pressures, which might be attributed to the smaller size of the propylene. Indeed Raman spectroscopy measurements performed for both propane and propylene (Figures S5 and S6 in supplementary information) reveal that both gases have identical effects on the framework, similar to effects caused by ethane adsorption. This result indicates that the π orbitals of the propylene have a minimal role in the interaction inducing the gate opening process.

An additional parameter to consider when explaining the dependence of the gate opening pressure dependence on the chain length is the isosteric heat of adsorption (Figure S9, section 6 in supplementary information). The results show that the interaction energy is larger for longer than shorter chain molecules. Both the chain length and the interaction energy contribute to the gate opening phenomenon. It is worth noting that the hysteresis is stronger for longer molecules, an effect that can be attributed to the difference in their binding energies. The structural transformation is triggered by the formation of a hydrogen bond between the CH with the C=O bond of the bpdc ligand. Importantly, the C=O bond of a neighboring bpdc ligand in the 2D layers is also affected by the presence of the adsorbate. Therefore, the longer the chain of the guest hydrocarbon, the more likely both C=O bonds in the 2D layers on neighboring bpdc ligands are affected by this interaction and the lower the pressure opening point, as shown in Figures 1 and 3.

To test the dependence of the gate opening pressure on the *strength* of the hydrogen bonds, we examined the behavior of acetylene because its C-H terminal bonds are more acidic, and therefore expected to have a stronger H-bonding as shown by Hartmann *et al.*[18] Acetylene isotherms have indeed a different behavior from the other hydrocarbons as shown in Figures 3 (see bond lengths) and S1 c of supplementary information. The vdW-DF calculations confirm that there is a strong H-bond between C-H acetylene terminal group and C=O (shorter distance between the C-H and the C=O), although the interaction energy ~ -52.0 kJ mol$^{-1}$ is similar to that of ethane (Figure 3e). The difference can also be observed in the Raman spectra for acetylene adsorption (Figure 4) that are dramatically different from the Raman spectra of the other hydrocarbons. Moreover, a new strong mode at 1616

cm$^{-1}$ emerges and is accompanied by the disappearance of the C=O mode and the C≡C stretch mode in the bpee ligand is red shifted by ~-2 cm$^{-1}$. This 1616 cm$^{-1}$ mode, assigned to the C=O mode, is strongly hydrogen bonded to the acidic C-H terminal group of $C_2H_2$. This mode is similar to the mode observed for the as-synthesized RPM3-Zn, and can be attributed to the C=O bond when it is strongly H-bonded to the C-H of the DMF (Figure S8 in the Supplementary information). The blue spectrum shown in the bottom panel of Figure 4 shows an increase in ν(C-C) mode of the bpdc ligand and an increase in the in-plane deformation modes of the pyridine ring. The calculations show (Figure 4) that the angle between the two rings in the bpdc linker increases (Δφ = 2.33 degree) in agreement with the observed intensity increase of the ν(C-C) inter-ring stretch mode of the bpdc linker. Moreover, the smaller kinetic diameter of the acetylene (3.3 Å) as compared to the ethane (4.4 Å) also facilitates the gate opening process.

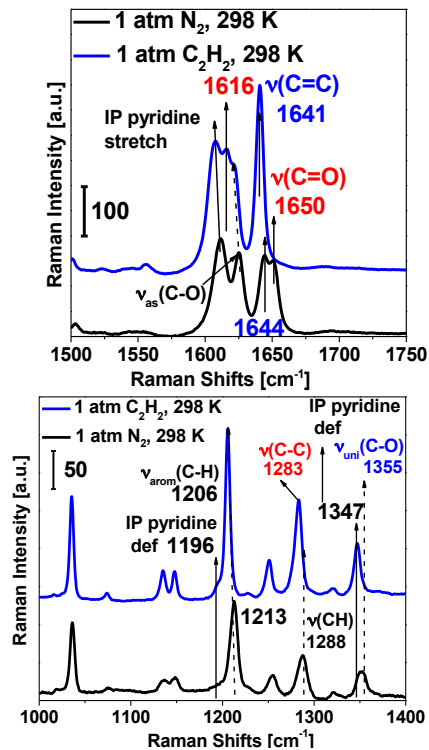

Figure 4. Raman spectra of RPM3-Zn under 1 atmosphere of $N_2$ (in black) and after introducing 1 atmosphere of acetylene (in blue) at room temperature. The top and bottom panels show the spectra in different frequency ranges.

vdW-DF calculations also confirm that acetylene, ethylene and ethane all strongly interact with the monodentate C=O bond of the bpdc ligand. On the other hand, their incorporation into the small pockets formed by "parallel" bpdc-units (Figure 3) is only ~ -1 kJ mol$^{-1}$ less favorable.

In summary, we present the first selective separation of C1-C4 paraffins and two pairs of C2 isomers ($C_2H_2$-$C_2H_4$ and $C_2H_4$-$C_2H_6$), in a flexible framework, RPM3-Zn, based on gas-framework interactions leading to differences in gate opening pressures. Raman spectroscopy and *ab initio* DFT calculations account for the separation behavior of the different hydrocarbons and show that H-bonding between their terminal groups and the C=O bond of bpdc ligand of the framework is the dominant effect. The separation behaviour of the C2 isomers is found to be dependent on the hydrogen bond strength and the presence of π electrons. The stronger interaction of longer chain hydrocarbons and the non-coordinated C=O bond in the bpdc ligands present in the 2D layers in RPM3-Zn, are key reasons for the gate opening pressure dependence (C2<C3<C4). Surprisingly, the effect of terminal

CH₃ groups is found to dominate over that of the high-density π-electrons in unsaturated hydrocarbons, accounting for the trends in gate opening pressures in similar size molecules. Strong H-bonding, as in the case of acetylene, reduces the gate opening pressure, although acetylene is smaller than the other hydrocarbons. This result confirms that stronger hydrogen bonding leads to a lower gate opening pressure and suggests a pressure swing adsorption type separation based on hydrogen bond strength in similar size molecules. This phenomenon also opens the door for the use of strong hydrogen bonding for the detection of traces of acetylene in gas flow for practical applications and as a sealing method for storage of adsorbed molecules in porous materials.

## ASSOCIATED CONTENT

### Supporting Information

Supporting information for the experimental details, adsorption-desorption isotherms of ethylene, ethane and acetylene, propane, propylene, X-ray diffraction data for 1-Hexene adsorption, Raman data for ethylene, propane and propylene, and RPM3-Zn before and after activation, the isosteric heat of adsorption for the differ alkane length, X-ray diffraction data, TGA, computational details, dihedral angle analysis and electron charge density and potential maps. This information is available free of charge via the Internet at http://pubs.acs.org/.

## AUTHOR INFORMATION

### Corresponding Author

Jingli@rutgers.edu, Chabal@utdallas.edu

### Author Contributions

‡These authors contributed equally. All authors contributed to the discussion and writing of the manuscript.

## ACKNOWLEDGMENT

This work was supported in its totality by the Department of Energy, Basic Energy Sciences, division of Materials Sciences and Engineering (DOE grant No. DE-FG02-08ER46491).

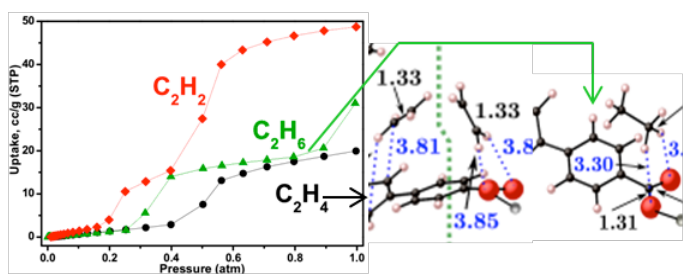